\documentclass[dvips]{article}
\usepackage{supertabular,lscape,epsfig}
\usepackage{amssymb}
\usepackage{amsmath}
\DeclareSymbolFont{ppa}{OT1}{ppl}{m}{it}
\DeclareMathSymbol{\vv}{\mathalpha}{ppa}{'166}

\begin{document}

\newcommand{\dd}{\,{\rm d}}
\newcommand{\ie}{{\it i.e.},\,}
\newcommand{\etal}{{\it et al.\ }}
\newcommand{\eg}{{\it e.g.},\,}
\newcommand{\cf}{{\it cf.\ }}
\newcommand{\vs}{{\it vs.\ }}
\newcommand{\zdot}{\makebox[0pt][l]{.}}
\newcommand{\up}[1]{\ifmmode^{\rm #1}\else$^{\rm #1}$\fi}
\newcommand{\dn}[1]{\ifmmode_{\rm #1}\else$_{\rm #1}$\fi}
\newcommand{\upd}{\up{d}}
\newcommand{\uph}{\up{h}}
\newcommand{\upm}{\up{m}}  
\newcommand{\ups}{\up{s}}
\newcommand{\arcd}{\ifmmode^{\circ}\else$^{\circ}$\fi}
\newcommand{\arcm}{\ifmmode{'}\else$'$\fi}
\newcommand{\arcs}{\ifmmode{''}\else$''$\fi}
\newcommand{\MS}{{\rm M}\ifmmode_{\odot}\else$_{\odot}$\fi}
\newcommand{\RS}{{\rm R}\ifmmode_{\odot}\else$_{\odot}$\fi}
\newcommand{\LS}{{\rm L}\ifmmode_{\odot}\else$_{\odot}$\fi}

\newcommand{\Abstract}[2]{{\footnotesize\begin{center}ABSTRACT\end{center}
\vspace{1mm}\par#1\par   
\noindent
{~}{\it #2}}}

\newcommand{\TabCap}[2]{\begin{center}\parbox[t]{#1}{\begin{center}
  \small {\spaceskip 2pt plus 1pt minus 1pt T a b l e}
  \refstepcounter{table}\thetable \\[2mm]
  \footnotesize #2 \end{center}}\end{center}}

\newcommand{\TableSep}[2]{\begin{table}[p]\vspace{#1}
\TabCap{#2}\end{table}}

\newcommand{\FigCap}[1]{\footnotesize\par\noindent Fig.\  %
  \refstepcounter{figure}\thefigure. #1\par}

\newcommand{\TableFont}{\footnotesize}
\newcommand{\TableFontIt}{\ttit}
\newcommand{\SetTableFont}[1]{\renewcommand{\TableFont}{#1}}

\newcommand{\MakeTable}[4]{\begin{table}[htb]\TabCap{#2}{#3}
  \begin{center} \TableFont \begin{tabular}{#1} #4
  \end{tabular}\end{center}\end{table}}

\newcommand{\MakeTableSep}[4]{\begin{table}[p]\TabCap{#2}{#3}
  \begin{center} \TableFont \begin{tabular}{#1} #4
  \end{tabular}\end{center}\end{table}}

\newenvironment{references}%
{
\footnotesize \frenchspacing
\renewcommand{\thesection}{}
\renewcommand{\in}{{\rm in }}
\renewcommand{\AA}{Astron.\ Astrophys.}
\newcommand{\AAS}{Astron.~Astrophys.~Suppl.~Ser.}
\newcommand{\ApJ}{Astrophys.\ J.}
\newcommand{\ApJS}{Astrophys.\ J.~Suppl.~Ser.}
\newcommand{\ApJL}{Astrophys.\ J.~Letters}
\newcommand{\AJ}{Astron.\ J.}
\newcommand{\IBVS}{IBVS}
\newcommand{\PASP}{P.A.S.P.}
\newcommand{\Acta}{Acta Astron.}
\newcommand{\MNRAS}{MNRAS}
\renewcommand{\and}{{\rm and }}
\section{{\rm REFERENCES}}
\sloppy \hyphenpenalty10000
\begin{list}{}{\leftmargin1cm\listparindent-1cm
\itemindent\listparindent\parsep0pt\itemsep0pt}}%
{\end{list}\vspace{2mm}}
 
\def\TYLDA{~}
\newlength{\DW}
\settowidth{\DW}{0}
\newcommand{\dw}{\hspace{\DW}}

\newcommand{\refitem}[5]{\item[]{#1} #2%
\def\REFARG{#3}\ifx\REFARG\TYLDA\else, {\it#3}\fi
\def\REFARG{#4}\ifx\REFARG\TYLDA\else, {\bf#4}\fi
\def\REFARG{#5}\ifx\REFARG\TYLDA\else, {#5}\fi.}

\newcommand{\Section}[1]{\section{#1}}
\newcommand{\Subsection}[1]{\subsection{#1}}
\newcommand{\Acknow}[1]{\par\vspace{5mm}{\bf Acknowledgments.} #1}
\pagestyle{myheadings}

\newfont{\bb}{ptmbi8t at 12pt}
\newcommand{\xrule}{\rule{0pt}{2.5ex}}  
\newcommand{\xxrule}{\rule[-1.8ex]{0pt}{4.5ex}}  
\def\thefootnote{\fnsymbol{footnote}}
\begin{center}

{\Large\bf
H$\alpha$ Imaging of X-ray Sources in Selected\\
Globular Clusters with the SOAR Telescope}
\vskip1cm
{\bf
Pawe{\l}~~P~i~e~t~r~u~k~o~w~i~c~z\\}
\vskip3mm
{
Departamento de Astronom\'ia y Astrof\'isica, Pontificia
Universidad Cat\'olica de Chile, Av. Vicu\~na MacKenna 4860,
Casilla 306, Santiago 22, Chile \\
Nicolaus Copernicus Astronomical Center,
ul. Bartycka 18, 00-716 Warszawa, Poland\\
e-mail: pietruk@astro.puc.cl\\
}
\end{center}

\Abstract{
We present results of a search for objects with H$\alpha$
excess, such as cataclysmic variables (CVs) and
chromospherically active binaries (ABs), as counterparts to X-ray
sources detected with Chandra satellite observatory in six
Galactic globular clusters (GCs): M4, M28, M30, M71, M80, NGC~6752.
Binary systems play a critical role in the evolution of GCs, serving
as an internal energy source countering the tendency of GC cores
to collapse. Theoretical studies predict dozens of CVs in the cores
of some GCs (\eg 130 for M28, 40 for M30). A number of such binaries
is also expected outside the core radius. However, few CVs are known
so far in GCs. Using subtraction technique applied to images
taken with the 4.1-m SOAR telescope we have found 27 objects
with H$\alpha$ excess in the field of the observed clusters,
of which nine are likely associated with the clusters.
Four are candidate CVs, four candidate ABs, one could be either a CV
or an AB. One H$\alpha$ object seems to be a background galaxy,
while other 17 detected objects are probably foreground or background stars.
}
{Stars: binaries: close -- Stars: activity -- Stars: chromospheres --
Stars: dwarf novae -- novae, cataclysmic variables -- Hertzsprung-Russell
(HR) and C-M diagrams -- globular clusters: individual: M4 (NGC~6121),
M28 (NGC~6626), M30 (NGC~7099), M71 (NGC~6838), M80 (NGC~6093), NGC~6752}

\Section{Introduction}

Globular clusters are excellent laboratories for the studies of
dynamical processes in dense stellar environments. It is well
established on theoretical grounds that close binary stars
drive the dynamical evolution of GCs (Stod\'o{\l}kiewicz 1986,
Hut \etal 1992). For example, the presence of the binary stars
can delay or halt the evolution of a cluster toward core collapse.
Frequent encounters between passing binaries and single
stars may produce tightly bound systems forming a variety of exotic
objects such as cataclysmic variables (CVs), quiescent low-mass X-ray
binaries (qLMXBs), and blue stragglers.

Cataclysmic variables are interacting binaries containing a main-sequence
or slightly evolved secondary star losing mass via Roche lobe overflow
onto a white dwarf primary. If the strength of the white dwarf's magnetic
field is weak ($B<10^5$~G) then an accretion disk forms. It is believed
that the disk thermal instability is the cause of repetitive outbursts
observed in some CVs called dwarf novae (DNe). Cataclysmic variables
in quiescence typically have absolute magnitudes $6<M_V<12$.
One of the characteristic observational features of disk CVs
in quiescence is the presence of hydrogen emission lines.
CVs have X-ray luminosities $L_{\rm X}<10^{33}$~erg~s$^{-1}$.
Quiescent CVs lie on or near cluster main sequences, $\gtrsim2$~mag
below the turn-off in optical color-magnitude diagrams (CMDs).

A recent, extensive photometric survey by Pietrukowicz \etal (2008) has shown
that DNe in GCs are very rare indeed. On the other hand, theoretical studies
predict dozens of CVs in GC cores (\eg Ivanova \etal 2006). For example,
we expect about 200, 130 and 40 CVs in the cores of 47~Tuc, M28 and M30,
respectively. A number of such binaries (mostly primordial) is also
expected outside the core radius or even the half-mass radius.
As an example, dwarf novae M5-V101 (Oosterhoff 1941)
and M22-CV2 (Pietrukowicz \etal 2005) are located about two half-mass radii
from the cluster centers.

Chromospherically and magnetically active binaries (ABs) have X-ray
luminosities similar to those of CVs. There are three types of AB
systems: detached binaries comprised of a main-sequence star and a giant or
a subgiant (RS~CVn systems), detached binaries comprised of two main-sequence
stars (BY~Dra systems), and contact binaries (W~UMa systems).
ABs typically can be found on or somewhat above the cluster main
sequence or along the subgiant branch in CMDs. Some blue stragglers are
observed as contact binaries and can be active. It is believed that,
in all these kinds of systems, the magnetic activity is a result
of differential rotation and convection in the companion stars
controlled by tidal forces and mass transfer between them.
Active chromospheres of the stars are dominated by strong H$\alpha$
emission line at a wavelength of $6562.81$~\AA.

In this paper we present the results of a search for objects with H$\alpha$
excess, mainly CVs and ABs, as counterparts to X-ray sources
in six selected globular clusters. Section~2 gives details
on the observations and data reductions. Section~3 describes
the detected H$\alpha$ objects in each of the clusters, including
tentative classification. Finally, Section~4 states our main conclusions.

\Section{Observations and Data Reductions}

The observations were carried out on the night of 2008 Aug 7 using
the SOAR Optical Imager (SOI) mounted on the 4.1-m Southern
Astrophysical Research (SOAR) telescope located on Cerro Pach\'on,
Chile. The imager consists of two $2048\times4096$ pixel CCDs
spaced 102 pixels apart along their long sides. Each of the CCDs
is read by two amplifiers. The total field of view of SOI is
$5\zdot\arcm24\times5\zdot\arcm24$. We used $2\times2$ binning,
which resulted in a resolution of 0.153~arcsec/pixel. To cover the whole
$8\zdot\arcm8\times8\zdot\arcm8$ Chandra field of view and to fill
the $7\zdot\arcs8$ gaps between the individual CCD images, six
(or in some cases only four) single exposures were taken
in each filter. The following three filters were used:
$B$ (central wavelength $\lambda_0=4185$~\AA,
fwhm $\Delta \lambda=1030$~\AA), $R$ ($\lambda_0=6437$~\AA,
$\Delta \lambda=1525$~\AA), and H$\alpha$ ($\lambda_0=6563$~\AA,
$\Delta \lambda=75$~\AA), with exposure times of 150~s, 200~s
and 500~s, respectively. We observed six Galactic globular
clusters: M4, M28, M30, M71, M80 and NGC~6752. Tables~1 and~2
give basic physical facts on the clusters, such as reddening,
distance modulus, and metallicity, while Table~3 presents information
on timing, airmasses, and seeing conditions during the observations.

All images were de-biased and flat-fielded using SOI Reduction
Scripts\footnote{The scripts were taken from http://khan.pa.msu.edu/www/SOI/}.
With the help of the {\it Difference Image Analysis Package}
(DIAPL)\footnote{The package is available at
http://users.camk.edu.pl/pych/DIAPL/}, which is a modified version of
DIA written by Wo\'zniak (2000), we subtracted $R$-band images from
images taken in the H$\alpha$ filter (or vice versa if an
H$\alpha$ image had a lower seeing). We used the ESO Online Digital Sky
Survey (DSS) to set equatorial coordinates in each image. This operation
was done with an accuracy better than $1\zdot\arcs0$. All X-ray regions
in H$\alpha-R$ residual images were inspected by eye.

For construction of color-magnitude diagrams we extracted profile
photometry using the DAOPHOT/ALLSTAR package (Stetson 1987).
The H$\alpha$ magnitudes were calibrated by adopting $R-{\rm H}\alpha=0$
for the bulk of the stars. For all CMDs the $B$ and $R$ magnitudes were
scaled to an exposure time of 1.0~s. Each diagram shows only stars from
a single image, read by one amplifier, containing the central
part of a cluster, plus overlaid H$\alpha$-excess objects detected
in the whole observed field of the cluster.

\begin{table}[htb]
\begin{center}
\caption{\small General information on analyzed globular clusters
(part 1 of 2).}
\vspace{0.4cm}
{\small
\begin{tabular}{cccccc}
\hline
NGC  & Other &          RA(2000)          &          Dec(2000)       &     $E(B-V)$      &     $(m-M)_V$    \\
     & name  &                            &                          &       [mag]       &       [mag]      \\
\hline
6093 &  M80  & $16\uph17\upm02\zdot\ups5$ & $-22\arcd58\arcm30\arcs$ & $0.17 \pm 0.03^a$ & $15.58 \pm 0.12^a$ \\
6121 &   M4  & $16\uph23\upm35\zdot\ups5$ & $-26\arcd31\arcm31\arcs$ & $0.33 \pm 0.01^b$ & $12.51 \pm 0.09^c$ \\
6626 &  M28  & $18\uph24\upm32\zdot\ups9$ & $-24\arcd52\arcm12\arcs$ & $0.42 \pm 0.02^d$ & $13.4 \div 13.5^d$ \\
6752 &       & $19\uph10\upm52\zdot\ups0$ & $-59\arcd59\arcm05\arcs$ & $0.046\pm0.005^e$ & $13.24 \pm 0.08^e$ \\
6838 &  M71  & $19\uph53\upm46\zdot\ups1$ & $+18\arcd46\arcm42\arcs$ & $0.28 \pm 0.06^f$ & $13.71 \pm 0.11^f$ \\
7099 &  M30  & $21\uph40\upm22\zdot\ups0$ & $-23\arcd10\arcm45\arcs$ & $0.06 \pm 0.02^g$ & $14.65 \pm 0.12^g$ \\
\hline
\end{tabular}}
\end{center}
\vspace*{-0.3cm}
{\footnotesize
The columns give respectively: identification names,
equatorial coordinates of the centers, reddening $E(B-V)$,
distance modulus $(m-M)_V$. Coordinates were taken from
Harris (1996), while other references are given below the table.\\
References: $^a$~Brocato \etal (1998), $^b$~Ivans \etal (1999),
$^c$~Richer \etal (1997), $^d$~Davidge \etal (1996), $^e$~Gratton \etal (2005),
$^f$~Grundahl \etal (2002), $^g$~Sandquist \etal (1999)}
\end{table}

\begin{table}[htb]
\begin{center}
\caption{\small General information on analyzed globular clusters
(part 2 of 2).}
\vspace{0.4cm}
{\small
\begin{tabular}{ccccccc}
\hline
NGC  & [Fe/H] & $r_{\rm c}$ & $r_{\rm h}$ &      log $\rho_0$     & $N_{\rm CV}$ & Remarks\\
     &        &     [']     &     [']     & [L$_\odot$ pc$^{-3}$] &              & \\
\hline
6093 & $-1.71 \pm 0.20^a$ & 0.15 & 0.65 & 4.76 & 169      & \\
6121 & $-1.17 \pm 0.31^b$ & 0.83 & 3.65 & 3.82 & ~~~~~4.8 & \\
6626 & $-1.37 \pm 0.03^c$ & 0.24 & 1.56 & 4.73 & 130      & \\
6752 & $-1.48 \pm 0.07^d$ & 0.17 & 2.34 & 4.91 & ~~51     & ccc \\
6838 & $-0.71 \pm 0.11^e$ & 0.63 & 1.65 & 3.04 & ~~39     & \\
7099 & $-2.01 \pm 0.09^f$ & 0.06 & 1.15 & 5.04 & ~~39     & ccc \\
\hline
\end{tabular}}
\end{center}
\vspace*{-0.3cm}
{\footnotesize
The columns give respectively: identification name,
metallicity [Fe/H], core radius $r_{\rm c}$, half-mass radius $r_{\rm h}$,
central density $\rho_0$, predicted number of CVs in cores
of the clusters. All values but metallicities (see table footnotes)
were taken from Harris (1996). The predicted number of CVs in the
cores of the clusters was scaled to $N_{\rm CV}=200$ for 47~Tucanae
according to formula, given by Pooley \etal (2003), that the
encounter rate $\Gamma \propto \rho _0^{1.5} r_{\rm c}^2$.\\
References: $^a$~Brocato \etal (1998), $^b$~Drake \etal (1994),
$^c$~Davidge \etal (1996), $^d$~Gratton \etal (2003),
$^e$~Grundahl \etal (2002), $^f$~Sandquist \etal (1999)\\
Remarks: ccc -- core-collapsed cluster}
\end{table}

\begin{table}[htb]
\begin{center}
\caption{\small Short log of SOAR observations on the night of 2008 Aug 7/8}
\vspace{0.4cm}
{\small
\begin{tabular}{cccc}
\hline
NGC  &       UT     &   Airmass  & Seeing in $R$  \\
     &    [hh:mm]   &            &   [$\arcs$]  \\
\hline
6093 & 23:12--01:02 & 1.02--1.05 & 0.56--0.79 \\
6121 & 01:12--03:01 & 1.05--1.31 & 0.63--0.84 \\
6626 & 03:22--05:03 & 1.06--1.33 & 0.68--1.09 \\
6838 & 05:16--06:15 & 1.78--2.24 & 0.87--0.98 \\
6752 & 06:36--07:38 & 1.51--1.79 & 0.80--1.10 \\
7099 & 07:51--08:54 & 1.23--1.53 & 0.78--0.88 \\
\hline
\end{tabular}}
\end{center}
\end{table}

\Section{Detected H$\alpha$ objects}

We have detected 27 objects with H$\alpha$ emission around X-ray sources
in six observed globular clusters. Table~4 gives coordinates of the
objects, identifications of their X-ray counterparts, and tentative
classification. Figures~1-2 show finding charts in the $R$ band, and
corresponding H$\alpha-R$ residual images for all objects.
Below we describe in detail all H$\alpha$ objects we have found
in each of the six observed clusters.

\begin{table}[htb]
\begin{center}
\caption{\small Coordinates, X-ray counterparts, and classification
of detected objects with H$\alpha$ excess. Objects not classified
are very likely foreground or background stars.}
\vspace{0.4cm}
{\small
\begin{tabular}{lccccc}
\hline
H$\alpha$ object & RA(2000.0) & Dec(2000.0) & Distance &  ID number   & Class. \\
                    &            &             & from the &   of X-ray   & \\
                    &            &             & center   &    source    & \\
\hline
NGC6093-1 & $16\uph16\upm46\zdot\ups87$ & $-22\arcd55\arcm09\zdot\arcs6$ & $7\zdot\arcm47$ & J161646.9œôø²-225509$^a$ & \\
NGC6093-2 & $16\uph17\upm06\zdot\ups14$ & $-22\arcd58\arcm35\zdot\arcs6$ & $1\zdot\arcm06$ & J161706.2œôø²-225835$^a$ & AB? \\
\hline
NGC6121-1 & $16\uph23\upm34\zdot\ups44$ & $-26\arcd32\arcm00\zdot\arcs2$ & $0\zdot\arcm56$ & CX13$^b$ & AB \\
NGC6121-2 & $16\uph23\upm36\zdot\ups94$ & $-26\arcd31\arcm44\zdot\arcs5$ & $0\zdot\arcm42$ & CX15$^b$ & \\
NGC6121-3 & $16\uph23\upm40\zdot\ups25$ & $-26\arcd29\arcm25\zdot\arcs6$ & $2\zdot\arcm40$ & CX23$^b$ & CV \\
NGC6121-4 & $16\uph23\upm42\zdot\ups68$ & $-26\arcd31\arcm45\zdot\arcs1$ & $1\zdot\arcm82$ & CX24$^b$ & CV \\
\hline
NGC6626-1 & $18\uph24\upm20\zdot\ups62$ & $-24\arcd51\arcm26\zdot\arcs9$ & $2\zdot\arcm23$ &  \#2$^c$ & AB \\
NGC6626-2 & $18\uph24\upm22\zdot\ups85$ & $-24\arcd52\arcm45\zdot\arcs7$ & $2\zdot\arcm49$ &  \#5$^c$ & \\
NGC6626-3 & $18\uph24\upm24\zdot\ups24$ & $-24\arcd51\arcm03\zdot\arcs4$ & $2\zdot\arcm46$ &  \#6$^c$ & \\
NGC6626-4 & $18\uph24\upm24\zdot\ups63$ & $-24\arcd53\arcm01\zdot\arcs1$ & $2\zdot\arcm08$ &  \#7$^c$ & AB/CV \\
NGC6626-5 & $18\uph24\upm25\zdot\ups21$ & $-24\arcd54\arcm06\zdot\arcs1$ & $2\zdot\arcm70$ &  \#8$^c$ & AB? \\
NGC6626-6 & $18\uph24\upm32\zdot\ups41$ & $-24\arcd53\arcm52\zdot\arcs2$ & $2\zdot\arcm37$ & \#20$^c$ & AB? \\
NGC6626-7 & $18\uph24\upm33\zdot\ups70$ & $-24\arcd52\arcm13\zdot\arcs6$ & $1\zdot\arcm67$ & \#32$^c$ & AB \\
NGC6626-8 & $18\uph24\upm36\zdot\ups61$ & $-24\arcd50\arcm19\zdot\arcs6$ & $2\zdot\arcm59$ & \#38$^c$ & \\
NGC6626-9 & $18\uph24\upm37\zdot\ups34$ & $-24\arcd51\arcm55\zdot\arcs2$ & $1\zdot\arcm14$ & \#39$^c$ & \\
NGC6626-10& $18\uph24\upm38\zdot\ups93$ & $-24\arcd51\arcm45\zdot\arcs7$ & $3\zdot\arcm17$ & \#40$^c$ & \\
NGC6626-11& $18\uph24\upm39\zdot\ups50$ & $-24\arcd50\arcm29\zdot\arcs7$ & $0\zdot\arcm46$ & \#43$^c$ & \\
NGC6626-12& $18\uph24\upm40\zdot\ups26$ & $-24\arcd51\arcm34\zdot\arcs5$ & $1\zdot\arcm96$ & \#44$^c$ & \\
NGC6626-13& $18\uph24\upm42\zdot\ups63$ & $-24\arcd52\arcm47\zdot\arcs3$ & $0\zdot\arcm20$ & \#46$^c$ & \\
\hline
NGC6752-1 & $19\uph10\upm56\zdot\ups26$ & $-59\arcd59\arcm37\zdot\arcs8$ & $1\zdot\arcm21$ &  CX2$^d$ & CV \\
\hline
NGC6838-1 & $19\uph53\upm42\zdot\ups65$ & $+18\arcd45\arcm47\zdot\arcs9$ & $1\zdot\arcm26$ &  s02$^e$ & \\
NGC6838-2 & $19\uph53\upm43\zdot\ups84$ & $+18\arcd44\arcm27\zdot\arcs0$ & $2\zdot\arcm32$ &  s48$^e$ & \\
NGC6838-3 & $19\uph53\upm44\zdot\ups56$ & $+18\arcd48\arcm19\zdot\arcs9$ & $1\zdot\arcm68$ &  s49$^e$ & \\
NGC6838-4 & $19\uph53\upm52\zdot\ups75$ & $+18\arcd47\arcm52\zdot\arcs2$ & $2\zdot\arcm02$ &  s57$^e$ & CV?\\
NGC6838-5 & $19\uph53\upm52\zdot\ups80$ & $+18\arcd46\arcm35\zdot\arcs0$ & $1\zdot\arcm68$ &  s28$^e$ & gal\\
\hline
NGC7099-1 & $21\uph40\upm16\zdot\ups30$ & $-23\arcd08\arcm20\zdot\arcs1$ & $2\zdot\arcm81$ & \#33$^f$ & CV \\
NGC7099-2 & $21\uph40\upm21\zdot\ups44$ & $-23\arcd06\arcm50\zdot\arcs6$ & $3\zdot\arcm90$ & \#46$^f$ & AB \\
\hline
\end{tabular}}
\end{center}
\vspace*{-0.3cm}
{\footnotesize References: $^a$~Heinke \etal (2003), $^b$~Bassa \etal (2004),
$^c$~Becker \etal (2003), $^d$~Pooley \etal (2002), $^e$~Elsner \etal (2008),
$^f$~Lugger \etal (2007)}
\end{table}

\begin{figure}[htb]
\centerline{\includegraphics[width=140mm]{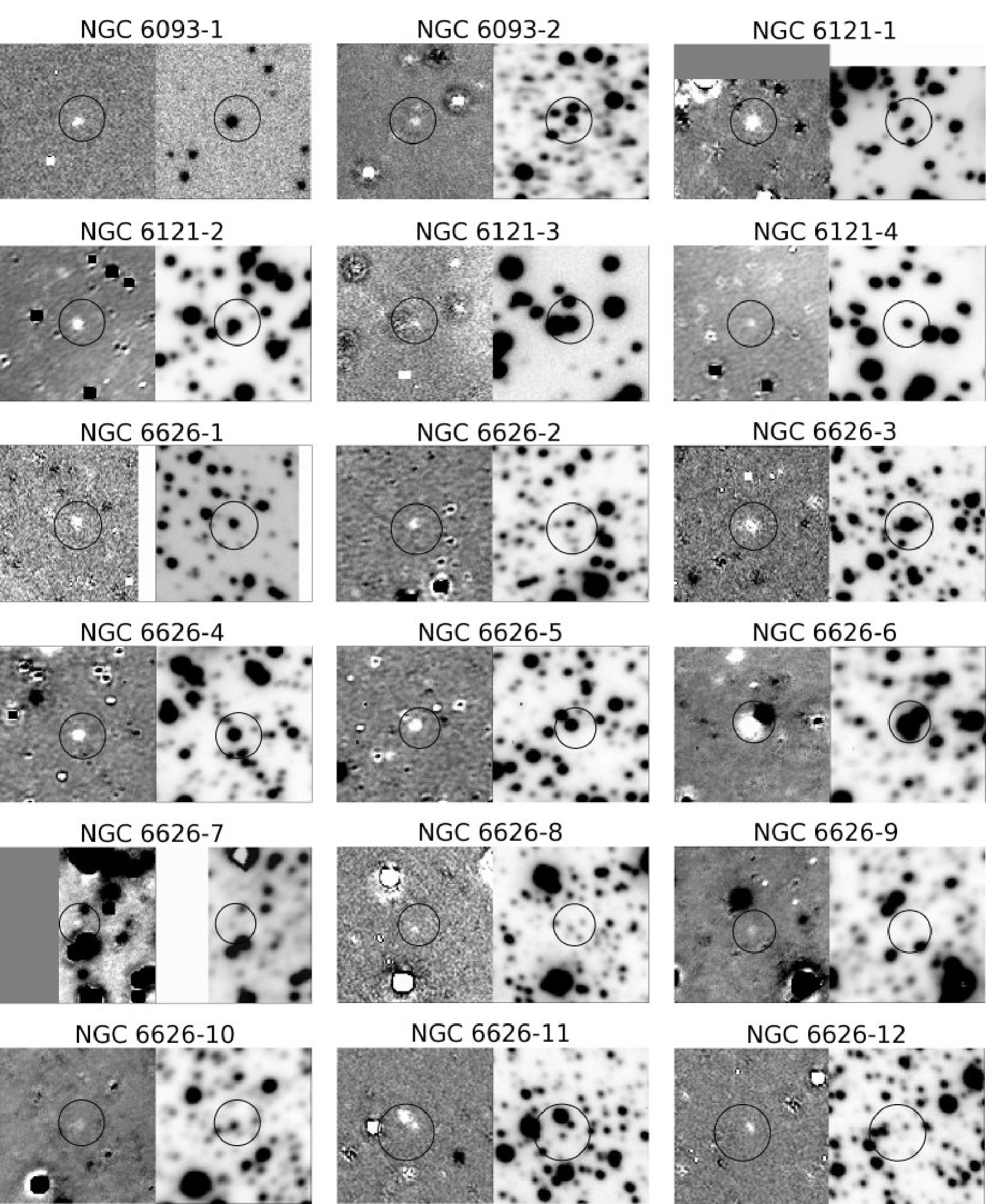}}
\vspace{0.4cm}
\FigCap{Charts for detected objects with H$\alpha$ excess
(part 1 of 2). The right-hand side panels show $R$-band images, while
the left-hand side panels show corresponding H$\alpha-R$ residual images.
The 99\% confidence uncertainties on X-ray positions are overlaid
on the charts. Each chart is $15\arcs$ on a side and centered on
an H$\alpha$ source. North is always up and East to the left.
}
\end{figure}

\begin{figure}[htb]
\centerline{\includegraphics[width=140mm]{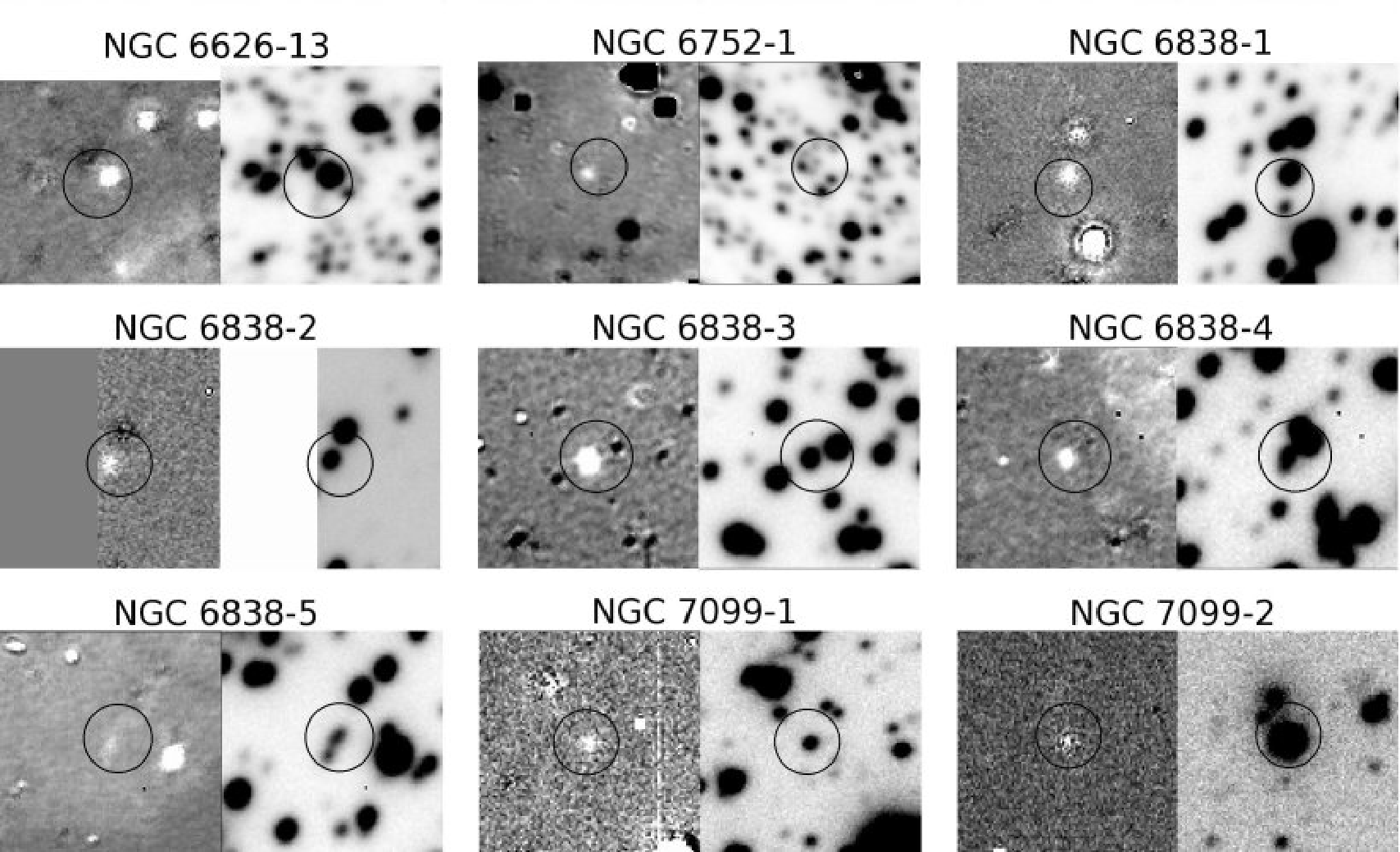}}
\vspace{0.4cm}
\FigCap{Charts for detected objects with H$\alpha$ excess
(part 2 of 2). The right-hand side panels show $R$-band images, while
the left-hand side panels show corresponding H$\alpha-R$ residual images.
The 99\% confidence uncertainties on X-ray positions are overlaid
on the charts. Each chart is $15\arcs$ on a side and centered on
an H$\alpha$ source. North is always up and East to the left.
}
\end{figure}

\Subsection{NGC 6093 (M80)}

This low-metallicity globular cluster has small core and
half-mass radii of only 9\arcs and 39\arcs, respectively.
NGC~6093 is unusual in having a variety of exotic objects.
It has one of the largest and most concentrated population of
blue stragglers (over 300) ever observed in a globular cluster
(Ferraro \etal 1999). NGC~6093 is known to harbor two dwarf novae
(DN1 and DN2, Shara \etal 2005) and a classical nova
(nova 1860 T~Sco, Shara and Drissen 1995). The cluster's
X-ray population is rich. Using Chandra satellite observatory
Heinke \etal (2003) detected some 19 sources within
the half-mass radius, and an additional 52 sources within
9\arcm~from the cluster center. Soft X-ray spectra of two bright inside
sources indicates that they are probable qLMXBs. Five sources
with hard spectra were classified as likely CVs. According to
theoretical calculations (see Table~2), we expect about
170 CVs in the core of NGC~6093. The reason for
such a large number of binaries in this cluster is its high
stellar density, and dynamical state near core-collapse.

We detected two H$\alpha$-emission objects around X-ray sources
in the field of NGC~6093. Finding charts are presented in Fig.~1.
Both of them lie outside the half-mass radius:
1.6$r_{\rm h}$ and 11.5$r_{\rm h}$ from the center.
The first one, object \#2, is faint, but has relatively large
H$\alpha$ excess (see Fig.~3). Although it was not possible to measure
its brightness in the $B$ band, this object is a good candidate
for an AB. Object \#1 is bright, but its location in the $R$ \vs $B-R$
diagram (Fig.3) and large angular distance from the cluster center
rather rule out this star as a member of NGC~6093.

The searches for counterparts to X-ray sources in this cluster
confirms that source 161714.6œôø²-225520 in Heinke \etal (2003)
coincides with the position of the bright star HD~146457 ($V=8.46$~mag),
which is saturated in our images. We note that the following X-ray sources
are located outside the analyzed field of view: J161648.5œôø²-225311,
J161701.0-œôø²225307, J161706.4-œôø²225318, J161712.3œôø²-225313,
J161713.6œôø²-225324, and J161721.7-œôø²225415.

\begin{figure}[htb]
\centerline{\includegraphics[height=90mm,width=130mm]{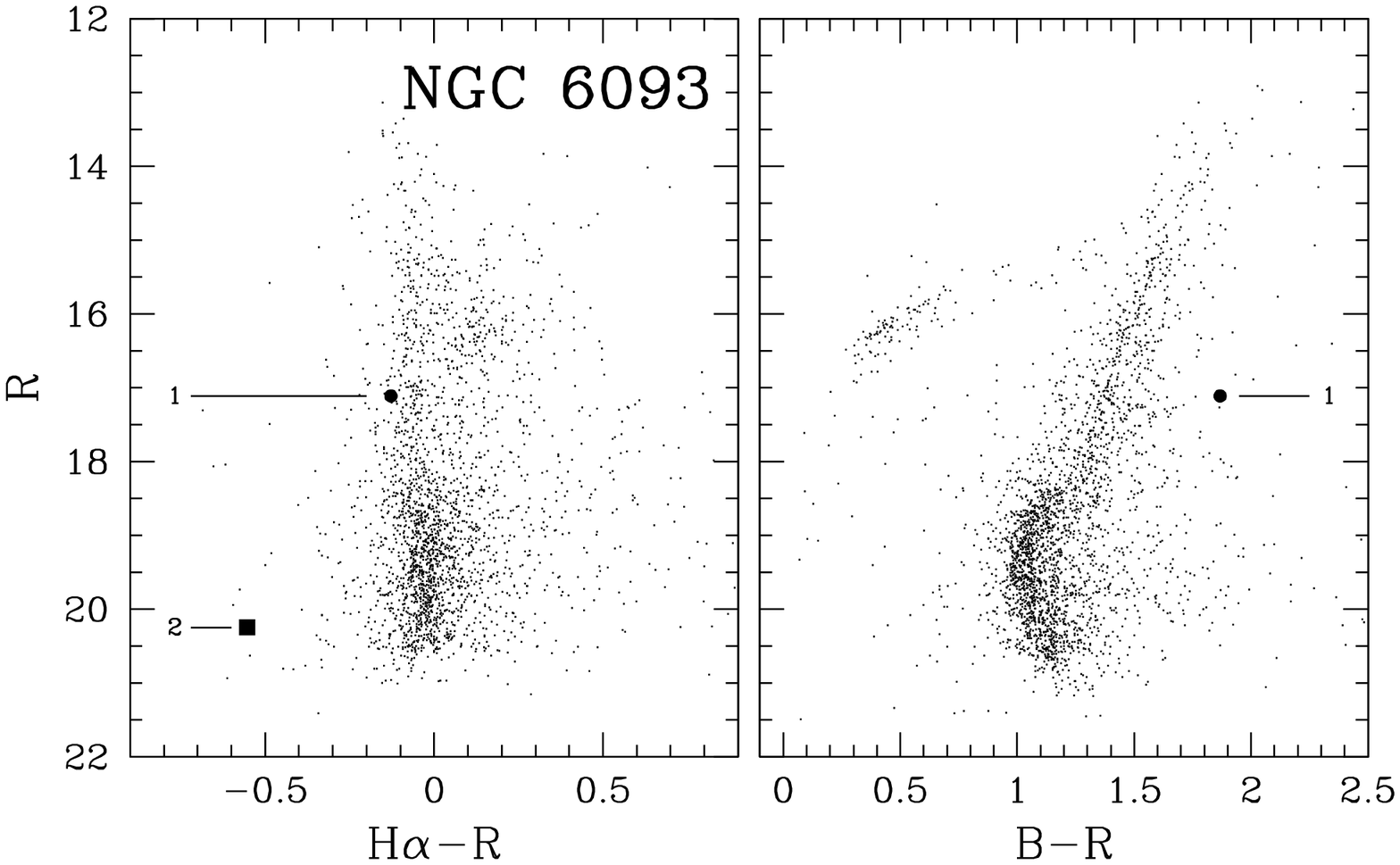}}
\FigCap{Color-magnitude diagrams for NGC~6093 (M80). Object \#2
could be a chromospherically active binary, though there is no
information on its $B-R$ color.}
\end{figure}

\Subsection{NGC 6121 (M4)}

NGC~6121 is the closest globular cluster (1.73 kpc, Richer \etal 1997).
The core and half-mass radii of this cluster are $0\zdot\arcm83$
and $3\zdot\arcm65$, respectively. NGC~6121 has a relatively low collision
number $\Gamma$, resulting from a relatively low central density $\rho_0$.
Thus the expected number of CVs in the core is not larger than five.
Bassa \etal (2004) report on detection of 31 X-ray sources with
luminosities down to $L_{\rm X}\approx10^{29}$~erg~s$^{-1}$.
According to the authors three sources (CX1, CX2, and CX4) are
probable CVs, because they have a relatively high X-ray-to-optical
flux ratio. One source (CX12) was classified as a millisecond
pulsar, and 12 sources as chromospherically active binaries.

The SOAR data allowed us to detect four stars with H$\alpha$
emission around X-ray sources (see charts in Fig.~1). All of them are
located inside the half-mass radius, while two stars, \#1 and \#2, even
inside the core radius. The two stars in the core are brighter
respectively by about 2~mag and 1~mag in $R$ than stars \#3 and \#4.
Figure~4 shows two color-magnitude diagrams with positions of the
four detected H$\alpha$ objects. Objects \#1 and \#2 coincide
respectively with W~UMa-type variables V49 and V48 from
Kaluzny \etal (1997). The star \#2 is a candidate blue straggler
and could be chromospherically active. The system \#1 is a likely
field star, since it is observed about 0.2~mag to the red
of the main sequence and 0.5~mag below turn-off in the $R$ \vs $B-R$ diagram.
The CMD location of star \#3 supports the hypothesis that this is a CV
in quiescence in NGC~6121. The position of star \#4 is
consistent with both the AB and CV interpretations, although the first
one seems to be more plausible.

\begin{figure}[htb]
\centerline{\includegraphics[height=90mm,width=130mm]{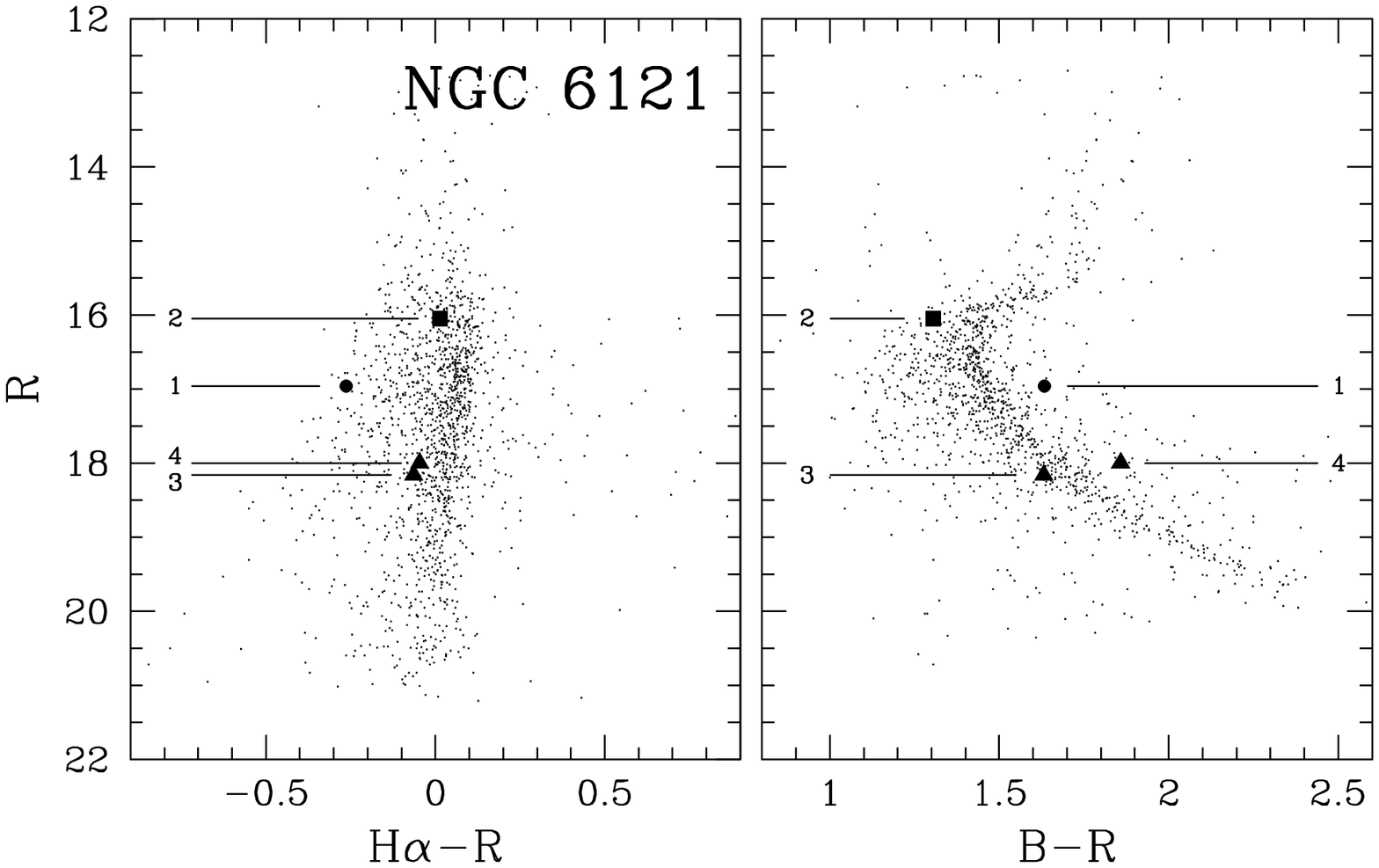}}
\FigCap{Color-magnitude diagrams with four H$\alpha$ objects
detected in NGC~6121 (M4). Object \#2, marked with a square, is an
AB candidate. Triangles denote the location of two possible CVs.}
\end{figure}

\Subsection{NGC 6626 (M28)}

NGC~6626 is located only $9\zdot\arcd6$ from the Galactic center
at ($l$,$b$)=($7\zdot\arcd80$,$-5\zdot\arcd58$), the most central
of all clusters studied here. It has a moderate reddening
of $E(B-V)=0.42$ and an average globular cluster metallicity
of [Fe/H]$=-1.37$ (Davidge \etal 1996). Chandra observations
of NGC~6626 allowed us to discover 46 X-ray sources, of which 12
lie within its $0\zdot\arcm24$ core radius (Becker \etal 2003).
Spectral analysis indicates that six of these sources are
neutron stars, including the pulsar PSR~B1821--24.

Our SOAR images have revealed 13 objects with H$\alpha$ excess
in the positions of X-ray sources in the field of this cluster.
Charts for these objects can be found in Figs.~1 and~2,
while color-magnitude diagrams are presented in Fig.~5.
Only three of the newly discovered H$\alpha$ objects,
namely \#1, \#4, and \#7, are located on the cluster
main-sequence or subgiant branch in the $R$ \vs $B-R$ diagram,
though it is not clear if source \#4 is a star.
Objects \#1 and \#7 are good candidates for RS~CVn systems
in NGC~6626. The other objects with H$\alpha$ excess
are likely foreground or background stars, including star \#13
which is located only 0.83$r_{\rm c}$ from the cluster center.

\begin{figure}[htb]
\centerline{\includegraphics[height=90mm,width=130mm]{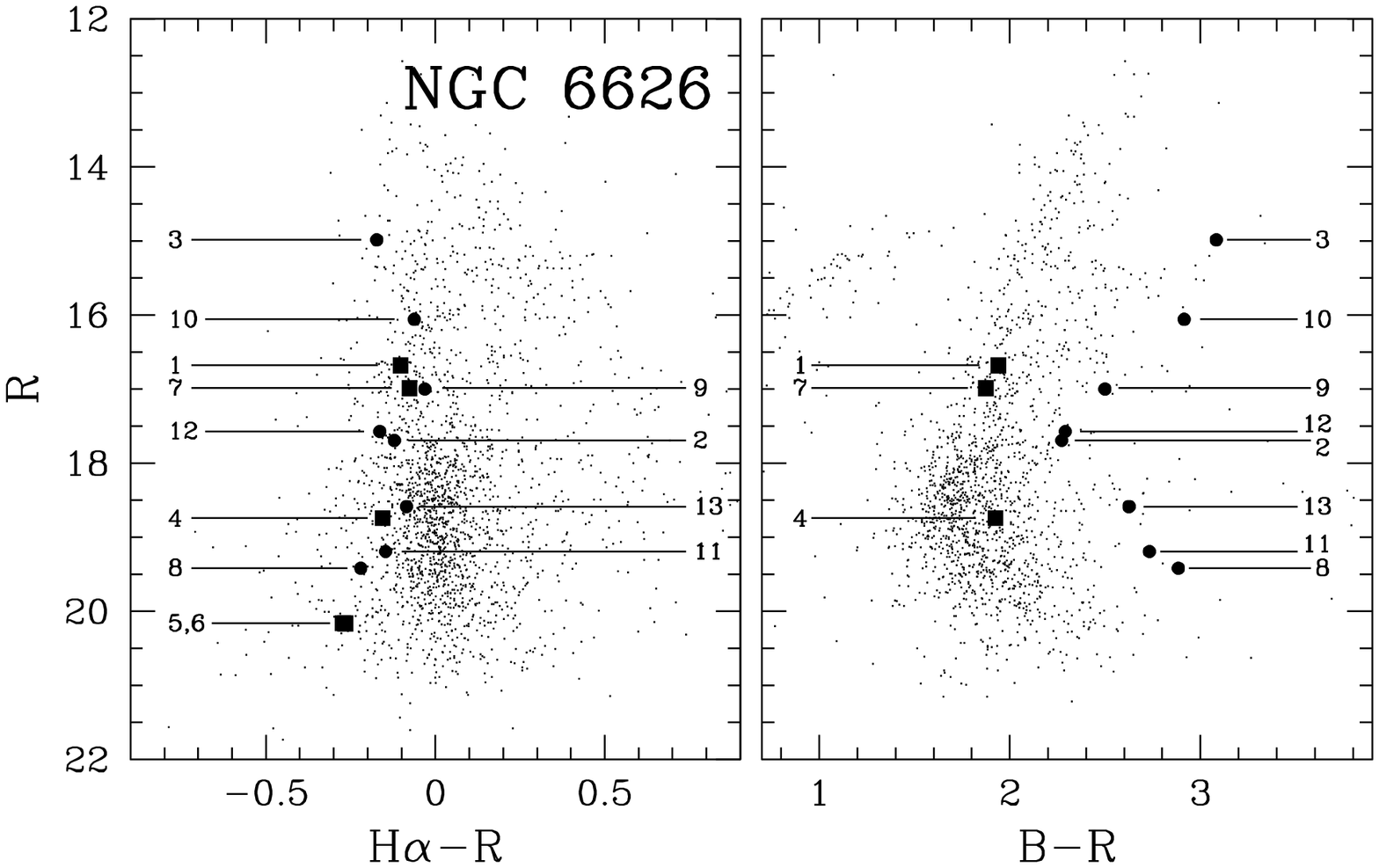}}
\FigCap{Color-magnitude diagrams for NGC~6626 (M28).
Objects marked with squares are AB candidates.}
\end{figure}

\Subsection{NGC 6752}

This core-collapsed globular cluster was observed with the Chandra X-ray
observatory on 2000 May 15. Pooley \etal (2002) report on detection
of 19 point sources within the cluster half-mass radius. Based on X-ray
and optical properties, they found 10 likely CVs, one to three likely
chromospherically active systems, and two possible extragalactic objects.
They detected another 21 sources outside the half-mass radius, but there is
no information available on these sources. Using archival H$\alpha$ images
of the central region of NGC~6752, taken with the WFPC2 camera on board HST,
Pooley \etal (2002) show that the optical counterparts to X-ray sources
CX1, CX2, CX4, and CX7 have clear H$\alpha$ excess. All four sources
were classified as CVs due to their characteristic X-ray spectra
(3~keV thermal bremsstrahlung).

In the SOAR data, only source CX2, located 1\zdot\arcm21
from the cluster center, has a noticeable H$\alpha$ emission.
The other three sources mentioned above are located at least three
times closer to the center, with CX4 and CX7 inside the core radius
of $0\zdot\arcm17$. For NGC~6752 we present only charts centered on
the optical counterpart of CX2 (see upper middle panel in Fig.~2).
Unfortunately, the star is not measurable in any of the SOAR images.

\Subsection{NGC 6838 (M71)}

This globular cluster is the most metal-rich ([Fe/H]$=-0.71$,
Grundahl \etal 2002) and has the closest location to the Galactic plane
($b=-4\zdot\arcd56$) among the six clusters observed with SOAR. The X-ray
population of NGC~6838 was recently studied by Elsner \etal (2008) with
the Chandra X-ray observatory. They found five sources located
within the $0\zdot\arcm63$ cluster core radius, another 24 sources
within the $1\zdot\arcm65$ half-mass radius, 34 sources with
$r_{\rm h}<r<2r_{\rm h}$, and 73 sources outside 2$r_{\rm h}$.
One of the five sources in the core is associated with the pulsar M71A.
For many X-ray sources located outside the core the authors
give possible optical and infrared counterparts in the 2MASS, USNO~B1.0
and TYCHO-2 catalogs. None of the sources is suspected to be
a cataclysmic variable.

Searches for H$\alpha$-emission objects in NGC~6838 with the SOAR telescope
have brought about five identifications. Finding charts for these objects are
included in Fig.~2, while $R$ \vs ${\rm H}\alpha-R$ and $R$ \vs $B-R$ CMDs
are presented in Fig.~6. All detections but object \#5 (associated
with the X-ray source s28) are of stellar nature. Star \#2 lies very close
to the edge of the CCD, and for this reason it is absent in both
color-magnitude diagrams. Stars \#1 and \#3 are located too far from
the cluster main sequence in the $R$ \vs $B-R$ diagram that their
membership status is out of question. The only object which
might be a CV is object \#4. Unfortunately it is surrounded by
bright stars, and no information on $B-R$ color was obtained.
Source \#5, due to its elongated shape, could be a starburst galaxy.

It is very likely that the X-ray source s52 is the bright ($V=10.76$~mag)
foreground star HD~350790. We also note that some of the X-ray sources
could not be checked, since they lie either inside the gaps between
the CCDs of the SOI camera (s34--s37) or outside the whole investigated
field of NGC~6838 (s61--s63, ss01--ss15, ss24, ss31, ss37--ss59).

\begin{figure}[htb]
\centerline{\includegraphics[height=90mm,width=130mm]{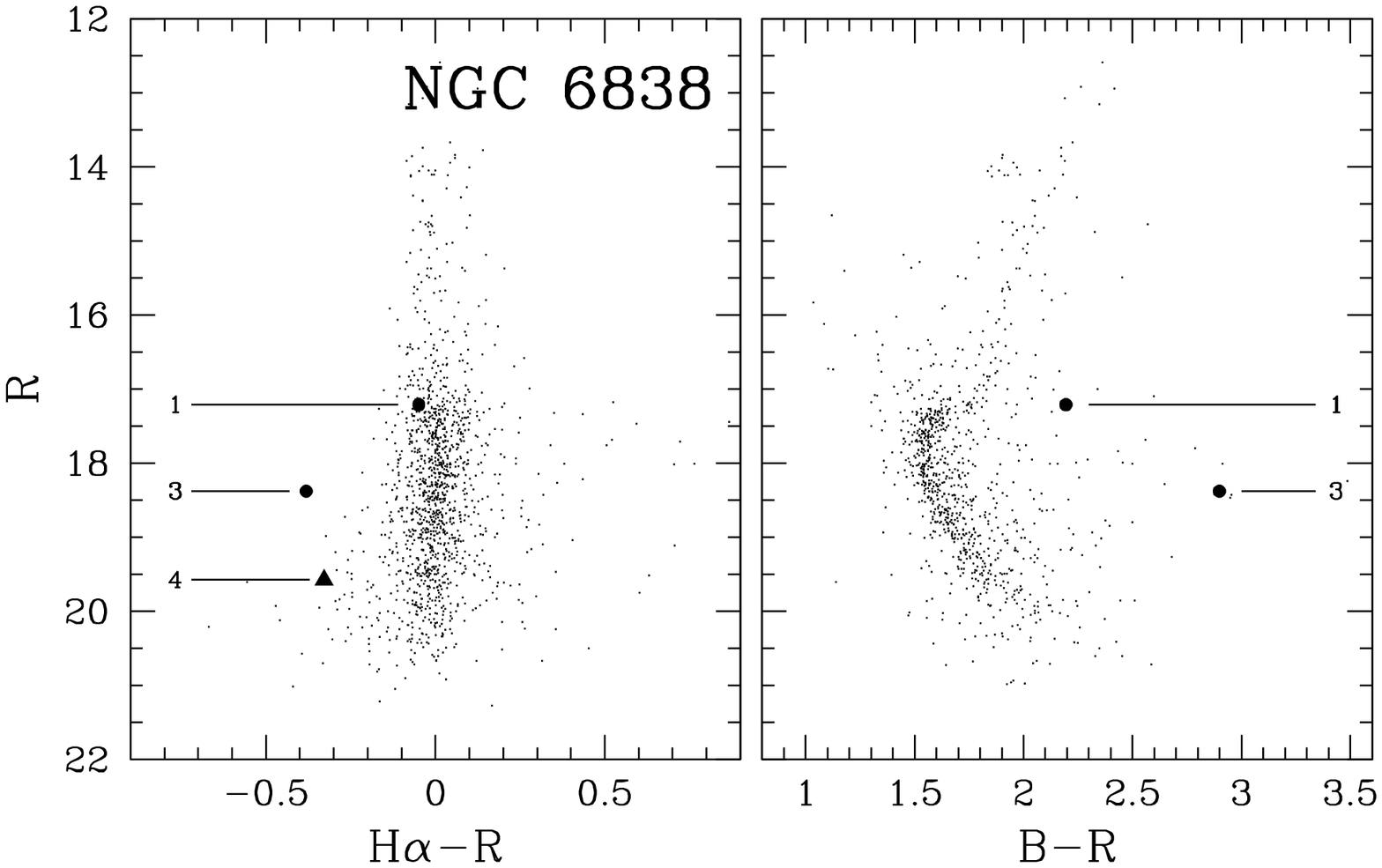}}
\FigCap{Color-magnitude diagrams for NGC 6838 (M71). The star marked
with a triangle could be a CV, but unfortunately there is no $B-R$ color
information for it.}
\end{figure}

\Subsection{NGC 7099 (M30)}

NGC~7099 has a low reddening of $E(B-V)=0.06\pm0.02$ and a distance
modulus of $(m-M)_V=14.65\pm0.12$ (Sandquist \etal 1999).
It is the second core-collapsed cluster in our sample. According
to Harris (1996), the central part of NGC~7099 represents
one of the highest density environments in the Galaxy. Thus,
a large population of compact binaries is expected in this cluster.
X-ray observations with Chandra (Lugger \etal 2007) revealed 50
sources within $4\zdot\arcm5$ from the center, with luminosities down
to $L_{\rm X}=4\times10^{30}$~erg~s$^{-1}$ at the distance of the cluster.
The brightest source ($L_{\rm X}=6\times10^{32}$~erg~s$^{-1}$) within
the $1\zdot\arcm15$ half-mass radius, source \#2 (also identified as A1),
is believed to be a qLMXB. Its blackbody-like soft X-ray spectrum supports
this hypothesis. Sources \#1 (A2), \#3 (A3), \#4 (B), and \#5 (C) have
X-ray properties consistent with being CVs. For sources \#7 and \#10,
Lugger \etal (2007) find possible AB counterparts in HST images.

The tiny and dense center of NGC~7099 is very difficult to study using
ground-based telescopes. The H$\alpha$ imaging with SOAR allowed us to
detect two sources: \#1 at a distance of 2.4$r_{\rm h}$, and \#2 at
3.4$r_{\rm h}$. The bottom of Fig.~2 includes charts for these two
stellar sources. Their location on the $R$ \vs $B-R$ diagram (Fig.~7)
favors the hypothesis that both objects belong to the cluster,
in spite of their relatively large distance from the center.
The fainter source \#1, which lies slightly to the blue side of the
main sequence $\sim$2~mag below of the turn-off, is a likely CV.
The CMD location of star \#2 is consistent with an AB interpretation.
Here, we also confirm that a bright foreground star at
$\alpha_{2000}=21\uph40\upm33\zdot\ups40$,
$\delta_{2000}=-23\arcd12\arcm36\zdot\arcs2$, which is
saturated in the SOAR images, is a counterpart to X-ray source \#39.
The positions of four X-ray sources, \#37, \#41, \#43, and \#50,
coincide with the positions of four galaxies in our images.

\begin{figure}[htb]
\centerline{\includegraphics[height=90mm,width=130mm]{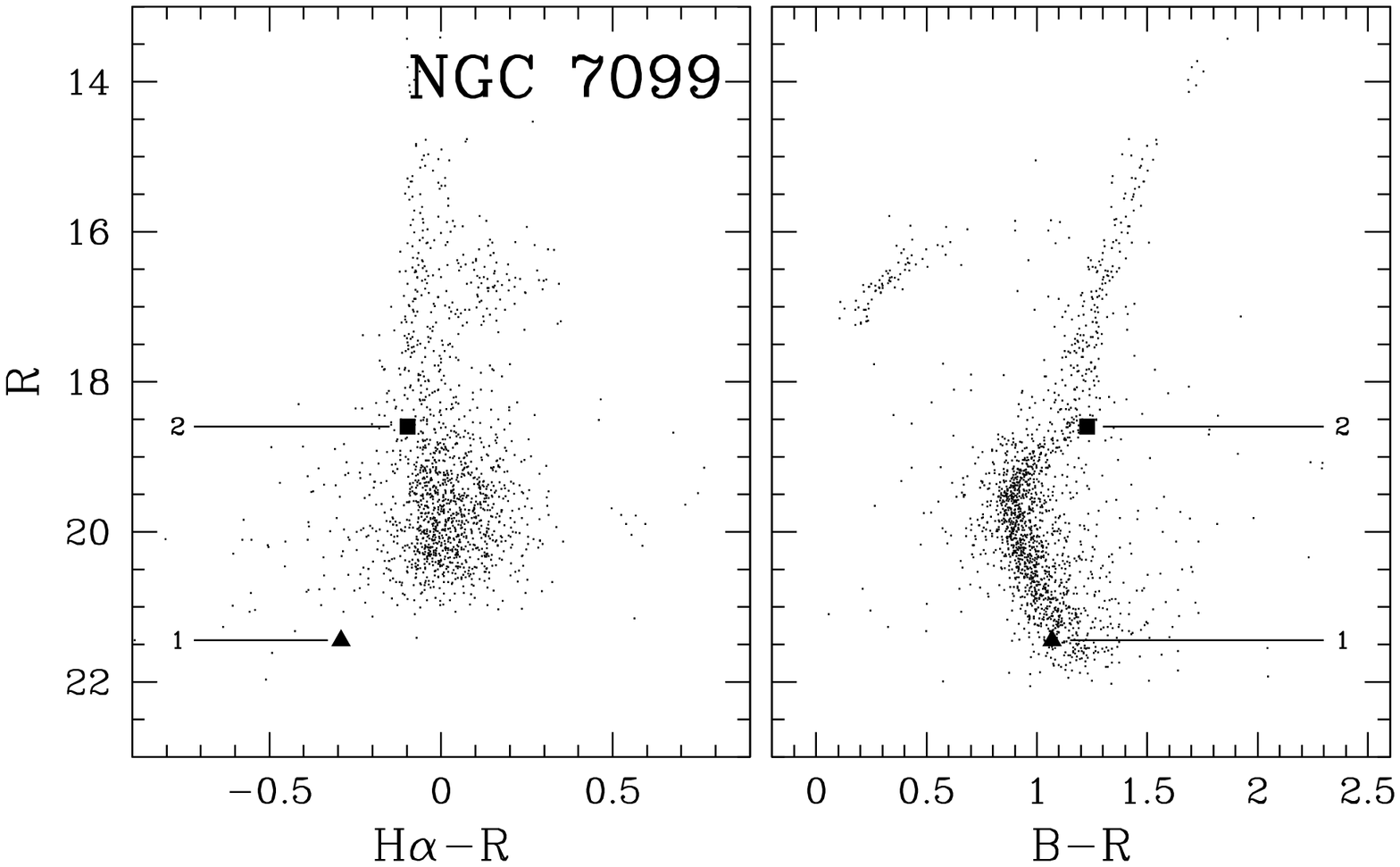}}
\FigCap{Optical color-magnitude diagrams for NGC~7099 (M30) with two
detected H$\alpha$} objects. Object \#1 is a candidate cataclysmic
variable, whereas object \#2 is a candidate for a chromospherically
active binary.
\end{figure}

\Section{Discussion and Summary}

We have presented the results of a search for objects with H$\alpha$ excess
in the field of six globular clusters: M4, M28, M30, M71, M80, and NGC~6752.
In total we have found 27 objects, of which nine seem to be associated with
these clusters. For four objects, due to lack of information on the $B-R$
color, their status membership remains unclear. For one source its elongated
image indicates an extragalactic nature. The rest of the detected H$\alpha$
objects are likely foreground or background stars.

The SOAR observations show how difficult it is to peer inside dense centers
of Galactic globular clusters using a medium-class, ground-based
telescope, even in good seeing conditions. Merely three
detected objects are located inside the cores of the investigated
clusters, and five more within the half-mass radii (see Fig.~8).
Source NGC6752-1 is the only object in the sample known to be a CV.
Unfortunately, it lies in a very crowded central region of this core-collapsed
globular cluster, and no brightness information could be obtained.
For nine objects likely associated with the clusters a tentative
classification is given based on their location on a $R$ \vs $B-R$
diagram. Four stars have been proposed to be chromospherically
active, four are possible CVs, and one object is either an AB or a CV.
Two of the four CVs, namely objects NGC6121-3 and NGC6121-4,
are located within the half-mass radius of M4, but none of them in the core.

The largest number of H$\alpha$-emission objects (13) has been found
in the field of the globular cluster NGC~6626 (M28). This cluster
is projected in front of the bulge, and it is not a surprise that
at least eight of these objects are foreground/background stars.

We stress that the SOAR observations were performed
on one night and that there are only single exposures for most
of the stars. Some of the detected H$\alpha$ objects are
variable, and the subtraction technique applied here can mimic
the H$\alpha$ excess. For example, it is not clear if the H$\alpha$
emission from stars NGC6121-1 and NGC6121-2, reported by
Kaluzny \etal (1997) as eclipsing/ellipsoidal binaries, is real.
A long-term monitoring would solve this problem.

The observed clusters have very different metallicities,
reddenings, and dynamical properties. Unfortunately, the small
number of detected objects is insufficient to show any tendency,
especially since for none of them membership status has been
confirmed. For all clusters but NGC~6121 the total number
of known CVs in their centers seems to be far too small
compared to theoretical predictions. However, one has to take into
account that H$\alpha$ emission from many faint ABs and CVs,
especially located in the crowded central parts of the clusters,
can be very weak or even undetectable. It is very likely that
new faint compact binaries such as CVs will be discovered in the near
future thanks to the newly refurbished Hubble Space Telescope.

\begin{figure}[htb]
\centerline{\includegraphics[width=140mm]{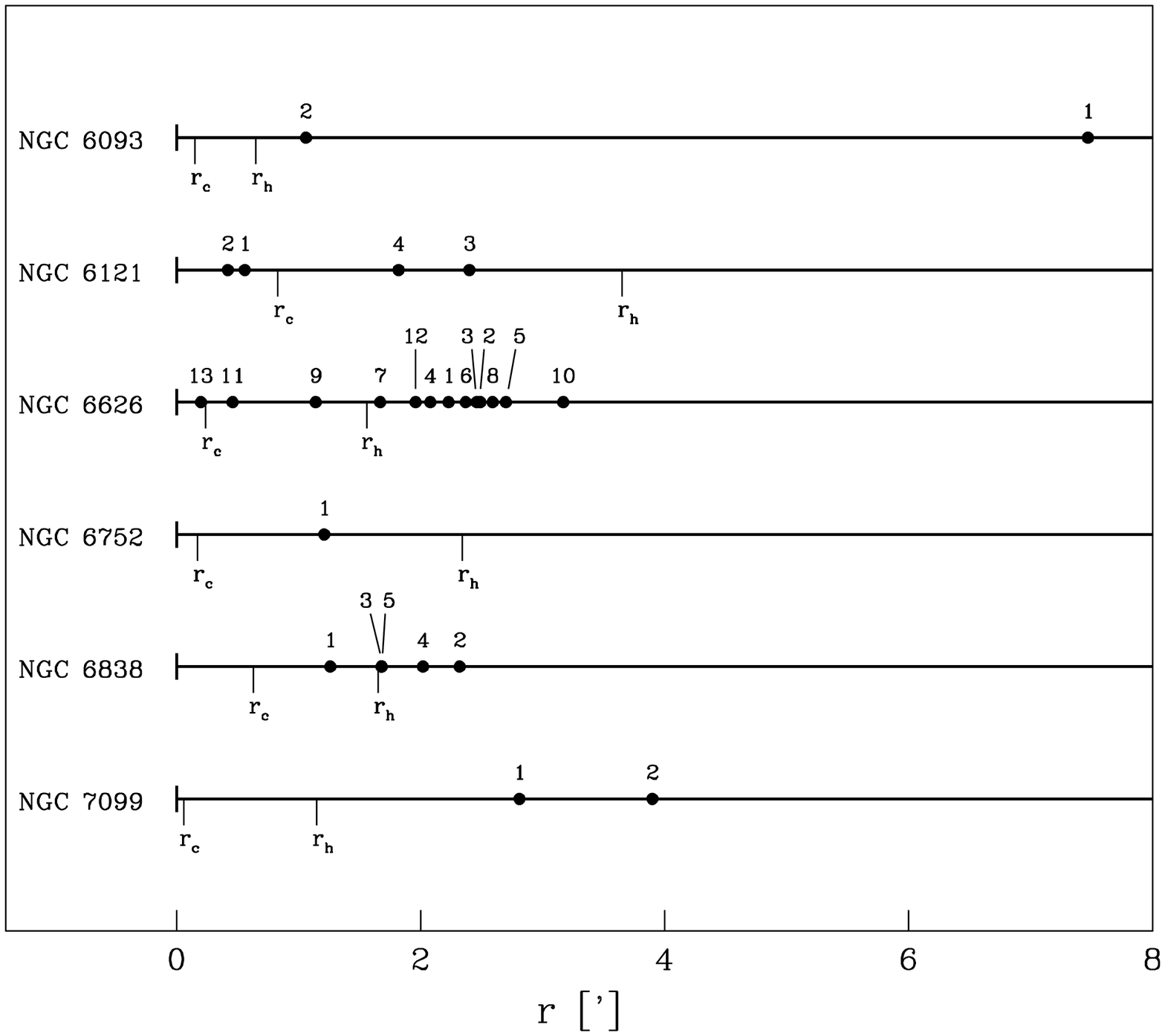}}
\FigCap{Distances from the cluster centers of the detected H$\alpha$
objects. Core and half-mass radii are marked for each of the clusters.}
\end{figure}

\Acknow{
The author would like to thank M\'arcio Catelan for important
discussions and remarks on the draft version of this paper.
The author is supported by the Chilean FONDAP Center for Astrophysics
No. 15010003, the Polish Ministry of Science and Higher Education
through the grant N N203 301335, and the Foundation for Polish
Science through program MISTRZ.
}

\end{document}